\documentclass[11pt,a4paper]{iopart}
\usepackage[dvips]{graphicx}

\begin{document}

\begin{center}
{\Large Cost and Capacity of Signaling in the 
{\it Escherichia coli} Protein Reaction Network}\\

\vspace{.5cm}

{\large Jacob Bock Axelsen$^{\dagger,*}$, Sandeep Krishna$^{*}$ and Kim
Sneppen$^{*}$\footnote{Corresponding author:
sneppen@nbi.dk}}\\

\end{center}

\vspace{.5cm}

{\small
\noindent
$\dagger$ Centro de Astrobiolog\'{\i}a\\ Instituto Nacional de T\'ecnica Aeroespacial\\Ctra de Ajalvir km 4, 28850 Torrej\'on de Ardoz\\
Madrid, Spain\\ 

\noindent
$*$ Center for Models of Life\\ Niels Bohr Institute\\ 
Blegdamsvej 17, 2100 \O\\ Copenhagen, Denmark}\\

\begin{abstract}
  
    In systems biology new ways are required to analyze the large amount
    of existing data on regulation of cellular processes.  Recent work
    can be roughly classified into either dynamical models of
    well-described subsystems, or coarse-grained descriptions of the
    topology of the molecular networks at the scale of the whole
    organism.  In order to bridge these two disparate approaches one
    needs to develop simplified descriptions of dy\-na\-mics and
    topological measures which address the propagation of signals in
    molecular networks.
    Transmission of a signal across a reaction node depends on the
    presence of other reactants. It will typically be more demanding
    to transmit a signal across a reaction node with more input
    links. Sending signals along a path with several subsequent
    reaction nodes also increases the constraints on the presence of
    other proteins in the overall network.  Therefore counting in and
    out links along reactions of a potential pathway can give insight
    into the signaling properties of a particular molecular network.

    Here, we consider the directed network of protein regulation in
    {\sl E. coli}, characterizing its modularity in terms of its
    potential to transmit signals.  We demonstrate that the simplest
    measure based on identifying sub-networks of strong components,
    with\-in which each node could send a signal to every other node,
    indeed partitions the network into functional modules.
    We suggest that the total number of reactants needed to send a
    signal between two nodes in the network can be considered as the
    {\em cost} associated to transmitting this signal. Similarly we
    define {\em spread} as the number of reaction products that could
    be influenced by transmission of a successful signal.
    Our considerations open for a new class of network measures that
    implicitly utilize the constrained repertoire of chemical
    modifications of any biological molecule.  The counting of cost
    and spread connects the topology of networks to the specificity of
    signaling across the network.
    Thereby, we address the signalling specificity within and between
    modules, and show that in the regulation of {\sl E.coli} there is
    a sys\-te\-ma\-tic reduction of the cost and spread for signals
    traveling over more than two intermediate reactions.
\end{abstract}

\maketitle

\section*{Background}

\noindent
Many functions of a living cell involve sending signals
from one protein to another. Signals need to be sent in response to
environmental conditions in order to trigger the appropriate
functional proteins needed at that time. For example, the presence of
food metabolites in the surroundings triggers signals from membrane
receptors to proteins involved in chemotaxis and metabolism required
to make the cell move toward and utilize the food; or a sudden change
in the temperature triggers signals to proteins which buffer the cell
against the shock. Many signalling pathways found in living cells have
been studied and modeled in great detail: the PTS sugar uptake
\cite{TB}, chemotaxis \cite{BBS,ASBL}, heat shock \cite{APS}, unfolded
protein response \cite{AS}, the p53 network \cite{TSJ}, NF-$\kappa$B
signalling \cite{HLSB,KJS} and the SOS response to DNA damage
\cite{Aksenov, KrishnaSOS}, just to name a few.  All the computations done by the
regulatory system of a cell are used to make sure the right signals
get sent at the right times to the right places.

Not much is known about the large-scale organization of protein
networks in the cell and the connection between their architectural
principles and the propagation of signals within them. This is the
subject of investigation in this paper. 

%
%
The different overall types of reactions we have in the network are:
\begin{itemize}
\item transcription, where activated/inhibited polymerase complexes interacts with a promoter and regulates the transcription of downstream open reading frames.
\item complex-formation, where a complex is created from either monomers or other complexes (RNA-polymerases and filaments).
\item activation/inhibition, where a protein (e.g. enzyme) is modified by another enzyme by the addition of an organic compound (e.g. phosphate and methyl). 
\item metabolic/enzymatic, where a protein reacts with one or more small molecule(s) (e.g. transport and cleavage).
\end{itemize}

The EcoCyc database contains all this information to the level of
water, ions, sugars, fatty acids, phosphate groups etc. Whereas we
include enzymatic reactions with metabolic output, we prune the
network by removing all metabolic nodes.

Our approach is to study a simplified dynamics of signal propagation
on an organism-wide network of proteins and reactions. By comparing
with appropriate randomized versions of the network we pinpoint
features of the design of the real network that influence signal
propagation.


We chose to study {\it Escherichia coli} because it is the most
studied pro\-ka\-ryote and, hence, its network of interactions and
reactions is most complete; several databases exist for the regulatory
and metabolic interactions in {\sl E. coli}
\cite{EcoCyc,KEGG,RegulonDB,EP}.  There are many ways to represent the
full known molecular network of {\sl E. coli}.  The standard method,
used in a number of studies of biological and social networks
\cite{Farkasetal,NG}, has been to use an undirected graph. Although
easily tractable, such a representation does lose a great deal of
information about the interactions.

A graph representation which, for the regulatory network of a living
organism, adds most of this missing information is one where the
network is described by a directed, bipartite graph.  Such a graph
has two types of nodes: protein nodes and reaction nodes (including
reversible and irreversible metabolic and complex-formation
reactions, as well as transcription reactions). In our
representation a modified, {\it e.g.}  phosphorylated, protein is
assigned a different node from the original protein. In addition,
complexes of proteins are also assigned their own nodes. Further,
the links have direction.  Fig. 1A shows such a representation of
the protein network of {\sl E. coli}.

Even more information is contained in a representation of the
network as a list of reactions. The list adds to the bipartite graph
information about which neighbours of a reactant node are reactants
and which are products. This reaction list and the directed
bipartite graph are the representations we focus on in this paper.
To study the signalling in these networks we introduce two
quantities which measure different aspects of signal
pro\-pa\-ga\-tion.
These measures are built on the fact that transmission of a signal
across a reaction node depends on the presence of other reactants. In
particular we will assume that transmission of a signal across
reaction nodes with more input links puts more constraints on the
status of other molecules in the network. A simple measure for
the complications associated with sending a signal along a given
pathway is to count the total number of in links or the total number of
out links of reaction nodes along the pathway.

Given a signal pathway from protein A to protein B, we can ask how
many other types of proteins are required to be present to allow the
signal to propagate all the way.  This we call the "cost" of the
path. Another quantity is the number of alternate branches, along
the path from A to B, that the signal could be broadcast on.  This
we call the "spread" of that path.  Quantifying such measures is
useful only if there is an appropriate null-model to compare with
the real {\sl E. coli} network. For this null-model we choose a
randomized version of the real network which has the same number of
nodes and links, which preserves bipartiteness as well as all local
point properties by keeping the in and out degree of
each node fixed.\\

\section*{Results}

\subsection*{Modular Design of the {\sl E. coli} Network}
The directed, bipartite graph representation of {\sl E. coli} consists
of 2846 protein nodes and 2774 reactions. The types of reactions are
transcription reactions, complex formations, protein modifications and
metabolic reactions. The dataset counts 848 transcription reactions
out of the 980 irreversible reactions, with the remaining 1794
reactions being reversible. In Fig.1A we show the giant
weak component consisting of 1938 reactions (of which 812 are
transcription reactions, (cyan squares)) and 1897 proteins (orange
circles). With such a network representation, one can identify four
different types of degree distributions: the in- and out-degree
distributions for protein and reaction nodes, shown in Fig
1C,D.


For the four different degree distributions only the out-degree
distribution of protein nodes is sufficiently broad to be fitted to a
power law with exponent of $\gamma=2.2$ over two decades; the other
three are narrow, exponential distributions.  The in- and out-degree
distributions of the reaction nodes reflect constraints on both space
and the number of constituents of each protein (or complex), with the
out-degree being slightly higher. The broadness of the out-degree
distribution of protein nodes is wholly due to transcription
reactions. Without these, the out-degree distribution of protein nodes
is almost indistinguishable from the in-degree distribution.

Another clear feature of the overall design is the tendency of
transcription reactions (cyan, in Fig. 1A) to be in the
center of the network. That is, if we simply count distances along
undirected paths starting from transcription reaction nodes we get an
average length of $\approx 4$. In contrast, the average length of paths
starting from arbitrary reaction nodes is $\approx 7$. This observation
is a rough approximation to what is captured by the betweenness
centrality measure\cite{Freeman}.

The alternating reaction and protein nodes as one moves away from the
core of the network in Fig. 1A is in part due to the bipartiteness and
in part due to the higher interconnectedness of the core of the
network, consisting mostly of transcription factors.  The average
degree of transcription factors is $\approx 11$, while it is $\approx
3$ for all proteins.

Fig. 1A illustrates that the {\sl E. coli} graph is
composed of a large number of relatively small strong components (a
strong component is a subgraph where there is a path between every
pair of nodes, see Methods section). The largest of these contains 150
nodes.  We will here refer to a graph where every node has access to
every other node through a path in the network as being above
percolation threshold or super-critical. Then, although the full
network shown in Fig. 1A looks supercritical, the
representation in terms of strong components shows that it is
substantially below the percolation threshold (as confirmed by the
exponential size distribution of strong components, not shown).  Fig.
1B shows a corresponding condensed graph of the randomized
network, in which the degree of each node is conserved.  The existence
of a giant strong component with $\approx 2000$ nodes (out of 3835 in
the giant weak component and 5620 in the full network) confirms that
there are enough links in the system to put it substantially above the
percolation threshold.  Thus, the known {\sl E.  coli} reaction
network indeed shows a highly modular design, even when compared to a
random bipartite network that has exactly the same number of nodes,
each with the same in- and out-degree.

\subsection*{Downstream Targets and Restrictions on Allowed Paths}
The simplest aspect of the structure of the network that influences
signalling is the number of nodes that are downstream of any given
starting node. Note that this is a quantity that can be sensibly
studied only with a directed graph representation of the network; in
any connected undirected graph all nodes are downstream of each other.
The possible signals emanating from the starting node are obviously
limited to reach only these nodes. The strong component structures in
Fig. 1A,B already indicate that the real {\sl E. coli}
network differs substantially from its randomized counterpart.  In the
random network most nodes can reach almost all other nodes, whereas
each protein in the real network has a much smaller number of
downstream targets.  Thus, the real network is relatively optimized
for specific signalling; a percolating structure is not conducive to
specific signalling because every node has almost the entire network
downstream of it.  This expectation is confirmed in Fig. 2A
which shows the distribution of the number of downstream targets for
the real and randomized {\sl E. coli} networks.

The fact that the {\sl E.coli} network has a few nodes with a
downstream sphere of influence of over 1000 indicates a topology
governed partly by a hierarchical subnetwork consisting of about 1/4
of the original network, as also noted by ref. \cite{MBZ}. In
contrast, the randomized network examined in Fig. 2A lacks such a
hierarchical organization, rather placing $\approx 2000$ nodes under
command of each other in one giant strong component.  Both of these
downstream spheres of influence are, however, subject to further
constraints.  Not all reactants in a reaction in fact provide a real
possibility to send a signal to each other. For example, a catalyst
can typically not receive a signal from any of the other reaction
partners.  We now investigate how such a constraint will affect
signalling in the {\sl E.coli} network.

Fig. 2B illustrates the kind of restrictions placed on
allowable signalling paths in a reversible reaction $A+B
\leftrightarrow C$. The graph representation does not have information
about these restrictions because all neighbors of a reaction node are
equivalent.  Including this restriction limits the downstream targets
from any node as compared to the simpler graph representation. This is
illustrated in Fig. 2C which shows the distribution of the
number of downstream nodes reachable from every node of the network in
Fig. 1 with the restrictions, as compared to Fig.
2A where the restrictions are not applied. Intriguingly, the
distribution with the signalling restrictions resembles a scale free
distribution, $1/n^{1.8}$, with a substantially better scaling than
the unrestricted signalling.  Irrespective of restrictions the real
{\sl E. coli} network has much less downstream targets than its
randomized version, a fact that is important for specific signalling.

\subsection*{Cost and Spread of a Path}
Signalling is not just about reaching a downstream target.  As a
signal propagates it needs other molecules to help it pass the message
across consecutive reactions.
Consider for example a signal initiated by an increase in the
concentration of a given transcription factor.  The promoter it
influences may depend on other transcription factors, for example in
an or-gate construction. If that is the case, and the other
transcription factor is already abundant, the promoter activity will
not be influenced and thus the signal will not be transmitted. More
generally, for each additional reactant along a reaction pathway,
signal propagation gets increasingly coupled to the overall state of
the molecules in the cell.  The more reactions in the path, and the
more reactants in each reaction, the more the conditions that need to
be met for propagation of the signal.

A concrete example of a signalling pathway is the Arc two component
regulatory system illustrated in Fig. 3A. A receptor
protein (ArcB) receives an external stimulus (here, lack of oxygen),
gets phosphorylated, and then undergoes a series of two reactions
where the phosphate group is shifted between residues in ArcB, such
that finally ArcBp can transfer the phosphate group to ArcA.
Subsequently, phosphorylated ArcA acts as a transcription factor for a
large number of genes including the {\sl sucA} gene emphasized in the
figure.  In terms of signal propagation, we follow the signal from a
phosphorylation reaction: $signal+ATP+ArcB \leftrightarrow ArcBp$,
through the reaction $ArcBp+ArcA \leftrightarrow ArcAp+ArcB$, ending
in the reaction $ArcAp+IHF+Fnr+RNAP\sigma^{70} \rightarrow
SucABCD+..$.

The external signal propagates under the condition that
all reactions can take place.  This means that (1) ArcB is present,
(2) ArcA is present, and that (3) the three additional transcription
factors (IHF, Fnr, and RNAP-$\sigma^{70}$) are present/absent in a
combination that allows a change in the concentration of ArcAp to
influence the activity of the {\sl sucABCD} operon. Thus, the
propagation of the input stimulus to SucA puts constraints on the
concentration levels of ArcA, ArcB, IHF, Fnr and the RNAP$\sigma^{70}$
complex, and can be assigned a cost ${\cal C}=5$ which counts the
number of proteins or protein complexes involved in propagating the
signal. In addition there could be some cost associated to the
absence/presence of small molecules or metabolites, for example ATP in
the first reaction of Fig.  3A.  We disregard this
metabolic part of signalling in the present paper.

We quantify this cost ${\cal C}={\cal C}({\rm path})$ for an arbitrary
path from a starting protein to a target protein by simply counting
the number of reactants along the entire path (not counting the
protein nodes which are part of the path), as described schematically
in Fig. 3B.  If the same reactant is used several times,
it is only counted once, as illustrated in Fig. 3C.
Notice that the propagation of a signal does not necessarily mean an
increased level of the proteins involved. The key point is that a
change in input state should be transmitted to a changed output state
of the end product.  Our cost function is a simple measure of the
complexity of handling such a signal and it could, in principle, be
calculated between any pair of proteins where a path exists in the
directed network.

Another issue which is important for specific signalling is the
possibility of signals branching, or spreading into the network. Thus,
a signal propagating from a starting protein to a target protein would
pass by some reactions where it could branch out into alternate paths
to different targets. Similar to the cost, we quantify this spread
${\cal S}={\cal S}({\rm path})$ for a given path from start to target
by counting the number of by-products along the entire path (Fig.
3B).  ${\cal S}$ does not count the sequence of products
needed to generate our final target, but only counts side-branches
along the path.

We stress that we here limit our spread counting to
reaction products (proteins) along the path,
whereas we disregard out links from proteins on the path that feed
into reactions. In principle these neighbor reactions to the path in
turn feed into changes of other proteins. Our minimal spread for
example disregard out degrees of highly connected transcription
factors along the path. This may sometimes be to restrictive,
but reflect the conjecture that specific disturbances typically
diminishes across a reaction node.
To be more specific on this last point, consider the case of a
transcription reaction where
the product $p=1/(1+r)$ as function of reactant $r$.
Here $p$ is only sensitive to $r$ when this is close to the
characteristic binding (here set to 1).
Thus for most values of $r$ the output response
$\delta p$ will be smaller than input changes $\Delta r$ across a
reaction node. For a related discussion on propagation of disturbances
in chemical reactions, see ~\cite{Maslov07}.

Fig. 4B shows the average cost of signals propagating
from one protein to another along the shortest path connecting them,
as a function of the length $l$ of that path. Each data point is the
average over all pairs which are at the given distance. Except for
paths of length two, the average cost for signals is significantly
smaller for the real {\sl E. coli} network than for a randomized
version which preserves degrees.  Fig. 4C shows the
average spread of signals propagating from one protein to another
along the shortest path connecting them, as a function of the length
of that path.  Each data point is the average over all pairs which are
at the given distance. As shown in Fig. 4A the number
of pairs at a given distance is quite high ($\sim 10^4$) for the real
network and much higher for the random. The standard error is
therefore negligible and not shown in Fig.4B,C. Just
as with the cost, except for paths of length two, the average spread
for signals is always significantly smaller for the real {\sl E. coli}
network than for a randomized version.

Notice that in the spread ${\cal S}$ vs. distance plot the slope, for
the random network, is $\Delta {\cal S}/\Delta l > 1$ whereas it is
$\Delta {\cal S}/\Delta l<1$ for the real {\sl E.coli} network. In
this connection keep in mind that a random directed network is
critical when the average out degree $\langle
k_{out}\rangle=2$. Considering a random path, a node on this path
should then on average have one more output than the one along the
path, corresponding to ${\cal S}=1$.
The values of $\Delta {\cal S}/\Delta$ then indicates that the
geometry of the random network is super-critical, with an initial
signal on average being amplified for each step along the path.  In
contrast the real network is sub-critical with signals that tend to
disappear with distance even under optimal conditions. Therefore, Fig.
1A,B can be regarded as a visual illustration of the
sub-criticality of the real network versus the super-criticality of
the randomized network.

In sum, the real {\sl E. coli} network reduces both the cost and
spread of signals along all shortest paths connecting pairs of
proteins.  Fig. 5 adds even more evidence to this conclusion by
showing that a scatter plot of spread vs. cost for all pairs of nodes
in the real {\sl E. coli} network covers a smaller area than a
corresponding plot for a randomized network. Note that this plot
contains the full distribution from whence the distance dependent
averages in Fig.4 were calculated.

Fig. 6 repeats this analysis for each of the six largest strong
components in the network. These strong components capture distinct
functional units being associated, respectively, to (a) predominantly
fatty acid metabolism, (b) the transcription network around $\sigma$
factors, (c) PTS-sugar transport, (d) ABC transporters, (e) the FeII
and FeIII transport system and finally, (f) the chemotaxis module.
Fig. 6 also shows the cost and spread for the constrained
reaction paths within each of these subgraphs compared to the expected
cost and spread for randomized versions of the subgraph.  Overall, we
see that cost and spread within each module is fairly similar to the
random expectation.
The only network which has a substantially lower cost and spread is
that of the ABC transporters, the network where signalling is most
seriously limited by the constraints.

\section*{Discussion and Conclusion}

We have shown that the molecular network of {\sl E. coli} is designed
in a way which optimizes signalling by minimizing its requirements on
the presence of other molecules, as well as focusing signalling on a
limited set of distant proteins with relatively small spreading of
signals to other proteins along the paths.  This overall design
feature is in accordance with the general belief that molecular
networks are somewhat modular \cite{Hartwell1999}.  Also this design
of the network consisting of relatively separated domains provides
much fewer alternate paths when compared to the random expectation.
Thus, the network is designed to favor specificity of signalling,
rather than provide robustness to deletion in the form of multiple
paths.  We take this as a hint that robustness is, presumably, a
design feature of the local dynamics in the network.  For example, the
well known robustness of chemotactic behavior is associated with
changes of reaction rates and protein concentrations \cite{ASBL}, but
not actual deletion of proteins.

We stress that our available network is based on literature study, and
therefore is vulnerable to systematic errors in collecting data.  In
particular, the overall data set probably covers only a fraction of
the real interactions in {\sl E.coli}. Further, certain types of
interactions are not available including, in particular, degradation
by proteases, RNA regulation and small molecule interactions.  Thus,
the observed sub-critical breakup of the network into separated strong
components in Fig. 1A may partly be due to limited data sampling. The
complete network of all interactions actually taking place in {\sl
E.coli} might well be above percolation.
This is especially likely to
be true if we also integrate the metabolism with the regulatory
network because much of the feedback in regulation goes through small
molecules involved in metabolic processes \cite{Sandeep}.


In regard to limitations of our approach to the incomplete {\sl E.coli} network,
it is important to emphasize that our measures of cost and
spread along a given path will be robust to improvement of the {\sl
E. coli} network.  The reason for this is that any reaction present in
the current network is well characterized, i.e., its set of reactants
and products is likely to be complete, and therefore its activity
should be fairly independent of presently unknown proteins. Thus,
improvement of the {\sl E. coli} network will likely involve addition
of new reaction pathways and will not, to a first approximation,
change the connections of the existing reaction nodes.  Therefore, for
any existing path in the current network the cost and spread will
remain unaffected. Adding further links to the network will increase
cost and spread for the random network, and thus tend to increase the
observed difference between signalling in the real and the randomized
network.

Looking at cost and spread within the strong components we found that
signalling within these modules was approximately as in their
randomized counterparts.  Thus, the cost and spread measure indeed
indicate a fair degree of robustness within a module, while still
showing a systematic absence of alternate path options on large
scales.  However, examining these modules against deletion of
individual nodes we found that, for all the six largest strong
components, the robustness of the size of the module was less than for
a comparable module with randomized structure.  Thus, even within
modules, percolation robustness of signals is not a strong trait.

It is clear that our definition of cost in terms of simply counting
independent inputs is a simplified approach. Thus, one could easily
imagine constructing more complicated cost functions, taking into
account, in particular, the logic of transcription regulation
\cite{HSWK,CSP} and epigenetic switches\cite{Mille}. Also the cost may
be modified according to universally abundant proteins (housekeeping
genes), for example by not counting input from all essential genes. To
some extent our counting already excludes core enzymes such as
ribosomes and tRNAs but, obviously, this list of essential ingredients
of cell functionality may be extended.  Finally, the real usage of a
given pathway may be restricted by the time to process the signal
along the path, wherein particular protein production events take a
sizable time compared to a cell generation.

A final intriguing point is that the large modules have such widely
different design features, as seen from Fig. 6.  Indeed, some modules
C,F are dominated by complex formation reactions, D,E by linear
pathways, while A,B are densely interconnected.  Thus, whereas
signalling within each of the sub-networks is similar to random, in
terms of cost and spread, the way these networks deal with the
signalling is still widely different. We could not detect motifs
common to all of these macromolecular networks~\cite{Alon}.


As an overall summary, our geometrical considerations capture
a modularity of the {\sl E.coli} protein networks which favors
signaling on fairly short distances: A topology which speaks to
fruitful modular approaches to systems biology on the whole-cell scale,
as propagation of signals through many intermediate reactions
seems to be nearly impossible.
In addition, one expects limitations in signal propagation
from simple mass-action kinetics, as shown by
[cite Sergei Maslov, Kim Sneppen, Iaroslav Ispolatov
``Propagation of fluctuations in interaction networks
governed by the law of mass action"
q-bio.MN/0611026].
As the macromolecular network in {\sl E.coli}
indeed has modular features, and signals are difficult to
transmit, substantial parts of {\sl E.coli} may be consistently
understood by summing up separate studies of nearly independent modules.

\section*{Methods}

\subsection*{Network construction}
The basic flat files of the EcoCyc database \cite{EcoCyc} were
downloaded from \verb+Ecocyc.org+.  EcoCyc is a scientific database
for the bacterium {\it Escherichia coli} K-12 MG1655. The EcoCyc
project performs literature-based curation of the entire genome, and
of transcriptional regulation, transport and metabolic
pathways.

%
%
Despite being incomplete in places, when compared to more specialized
databases, EcoCyc is still the most comprehensive database of
reactions in {\it Escherichia coli}.

The files \verb+proteins.dat+ and \verb+genes.dat+ contains the list
of all proteins and gene names in the EcoCyc. From the files
\verb+bindrxns.dat+ and \verb+promoters.dat+\ all protein-promoter interactions
where extracted. The file \verb+transunits.dat+ contains a list of
specific transcriptional units which was used to link proteins to
their downstream gene products. These reactions where labelled
according to the name of the actual promoter involved in the
process. There is at least one promoter for each transcription
reaction in the database.

The files \verb+reactions.dat+ contains a general list of all
biochemical reactions in the EcoCyc, and the file \verb+enzrxns.dat+
specifies which of these are enzymatic reactions and which enzyme is
involved. From these files all other reactions where extracted where
at least one protein is at least a reactant or product.

From the total set of irreversible reactions (including all
transcription reactions) we removed proteins from the product side
which also occur as a reactant in the same reaction. The reason is
that information is not transmitted from reactants to catalysts,
therefore we do not want such links in our final network.

The resulting reaction list is represented as two stoichiometric lists
(matrices), one for reactants and one for products (proteins involved
in reversible reactions are also partitioned into two sets with one
being arbitrarily picked for the "reactant" matrix) of 2774 reactions
and 2846 proteins.

\subsection*{Randomization}
We constructed randomized versions of the {\sl E. coli} network by
repeatedly swapping the targets of randomly selected pairs of links
\cite{MS}.  This automatically preserves the in- and out-degree of
each node. Further, by restricting the set of pairs of links for which
swapping was allowed we could preserve both the bipartiteness and the
character of the links.  For instance, links to irreversible reactions
were only swapped with links to other irreversible reactions, etc. In
this way each (ir)reversible reaction remains (ir)reversible in the
randomized version.

\subsection*{Strong components}
It is possible to uniquely partition the nodes of any directed graph
into a set of strong components, see Fig.1A, bottom left.
Within each component, there is a path from every node of that
component to every other node in the component.  We generate the
strong components by selecting an arbitrary node and finding the
intersection between the set of nodes lying upstream and downstream to
the selected node.  This intersection plus the selected node forms one
strong component.  This process is repeated until all nodes are placed
in a strong component.  If there is no overlap between downstream and
upstream sets for a given node, then, by definition, that node is the
sole member of its strong component.  The partitioning produced by
this method is, for a given graph, unique and independent of the order
in which the nodes are chosen.

The condensed graph corresponding to a given directed graph is one
where each node represents one strong component of the original graph.
There is a directed link from one node to another if, in the original
graph, there is a link from any node of the first strong component to
any node of the second. The condensed graph, by definition, cannot
have any loops.

Notice that this partitioning into strong components is only possible
if there is transitivity of paths, i.e., if there exists a path from
node A to B, and from node B to C, then this implies there is a path
from A to C. Transitivity is essential to construct non-overlapping
strong components. If we restrict the allowed paths as described in
Fig. 2 then this is no longer true and therefore
non-overlapping strong components, as defined, cannot be constructed.

\subsection*{Cost and spread}
When calculating the downstream distribution in Fig. 2(A) \& (C) we
use a standard depth-first-search: we keep track of visited nodes so
that if we reach a node again by a longer path then it need not be
searched for by alternative paths further downstream. This method does
not take into account the bipartiteness of the graph.

We calculated cost and spread using a modified depth-first search of
paths in the graph. When restrictions of the type discussed in Fig. 2
are added the standard method is no longer sufficient (because of the
graph-theoretical non-transitivity of paths in bipartite graphs) and
the only way to enumerate all the shortest distance paths is to
actually go over all paths, of all lengths. In general, this is too
computationally expensive and therefore we put an arbitrary upper
cutoff on the length of allowed paths.  This restricts us to looking
at only those pairs which are within this cutoff distance.  However,
in practice, we are able to use a large cutoff of 14 (which covers
over 90\% of the pairs in the real network, see Fig. 4A) therefore
this does not affect our conclusions.

\section*{Authors contributions}
All authors contributed equivally to the work reported in this paper.

\section*{Acknowledgements}
The authors wish to thank the Danish National Research Foundation for
funding through the Center for Models of Life at the NBI. KS and JBA
wish to thank The Lundbeck Foundation. JBA wishes to thank The
Fraenkel Foundation.\\\\


\newpage

\section*{Figures}
  \subsection*{Figure 1 - {\sl E. coli} protein reaction network.}
  (A, Left) The graph is the largest weak component of a bipartite
  network, consisting of proteins (orange circles) and reaction nodes
  (promoters (cyan squares), complex formations \& modifications
  (black squares)).  The two largest hubs, $\sigma^{70}$ and $CRP$,
  and their links, have been removed for ease of visualisation.  (A,
  bottom left) Illustration of the procedure of condensing a directed
  graph (see Methods). An arrow indicates that there is a path
  connecting the two strong components in the original graph; nodes
  correspond to strong components of minimum size two. (A, Right) The
  resulting condensed graph of the {\sl E. coli} network.  (B) The
  similarly condensed graph for a randomized version of the {\sl E.
    coli} network.  (C) The cumulative degree distribution of reaction
  nodes for the full graph in (A).  (D) The cumulative degree
  distribution of protein nodes.

  \subsection*{Figure 2 - Domains of influence}
  (A) The cumulative distribution of number of downstream targets $s$
  without restrictions on allowed paths. Green is the randomized
  network (null hypothesis) and blue is the real network, the latter
  yielding a powerlaw distribution. (B) Schematic showing the
  restrictions on allowed paths for graphs constructed from a reaction
  list. The graph shown corresponds to a single reversible reaction:
  $A + B \leftrightarrow C$.  In the graph there is a path from {\it
    e.g.} $B$ to $A$, but in the real biochemical reaction this path
  does not exist. In contrast, paths from $A$ to $C$, and $B$ to $C$,
  are allowed. (C) Distribution of downstream targets with
  restrictions on the allowed paths. Notice how the distribution is
  now better resolved on nodes with high influence {\it i.e.} high
  $s$.

  \subsection*{Figure 3 - Cost and spread of a path.}
  (A) The Arc two-component regulatory pathway. (B) Schematic showing
  how the "cost" and "spread" of a signalling path, $A \leftrightarrow
  F$, is measured.  In this case protein $B$ and $D$ are necessary,
  giving a cost ${\cal C} = 2$. The proteins $E, G$ and $H$ are
  produced as a side effect, hence the spread is ${\cal S} = 3$. (C)
  Schematic illustrating the concept that if a protein is necessary
  for more than one reaction along the path, we count it only once.
  Thus, the cost is reduced to ${\cal C} = 1$, as compared to (B).

  \subsection*{Figure 4 - Measurements of cost and spread}
  (A) Number of pairs at a given (shortest) distance for the {\sl E.
    coli} network (solid line) and its randomized version (dashed
    line).  (B) Cost of a signalling path as a function of its length
    for the real (solid) and randomized (dashed) {\sl E. coli}
    networks.  (C) Spread of a signalling path as a function of its
    length for the real (solid) and randomized (dashed) {\sl E. coli}
    networks.
%
%
The shaded region illustrates which values lead to the strong
    components breaking up (if the network was infinitely large).

  \subsection*{Figure 5 - Scatter of cost {\it vs.} spread}
  Scatter plot of spread {\it vs.} cost for each pair of nodes lying within
  a distance of 14 to each other for the real (solid circles) and
  randomized (open circles) {\sl E. coli} networks.

  \subsection*{Figure 6 - The largest strong components}
  The six largest strong components of the {\sl E. coli} network,
  along with plots of the average cost, $\overline{{\cal C}(l)}$, and
  average spread, $\overline{{\cal S}(l)}$, as functions of signalling
  distance.  The yellow areas show the range spanned by
  $\overline{{\cal C}(l)}$ and $\overline{{\cal S}(l)}$ for 100
  randomized versions of the subgraphs.

\newpage

\textbf{Figure 1}\\

\includegraphics[width=\textwidth]{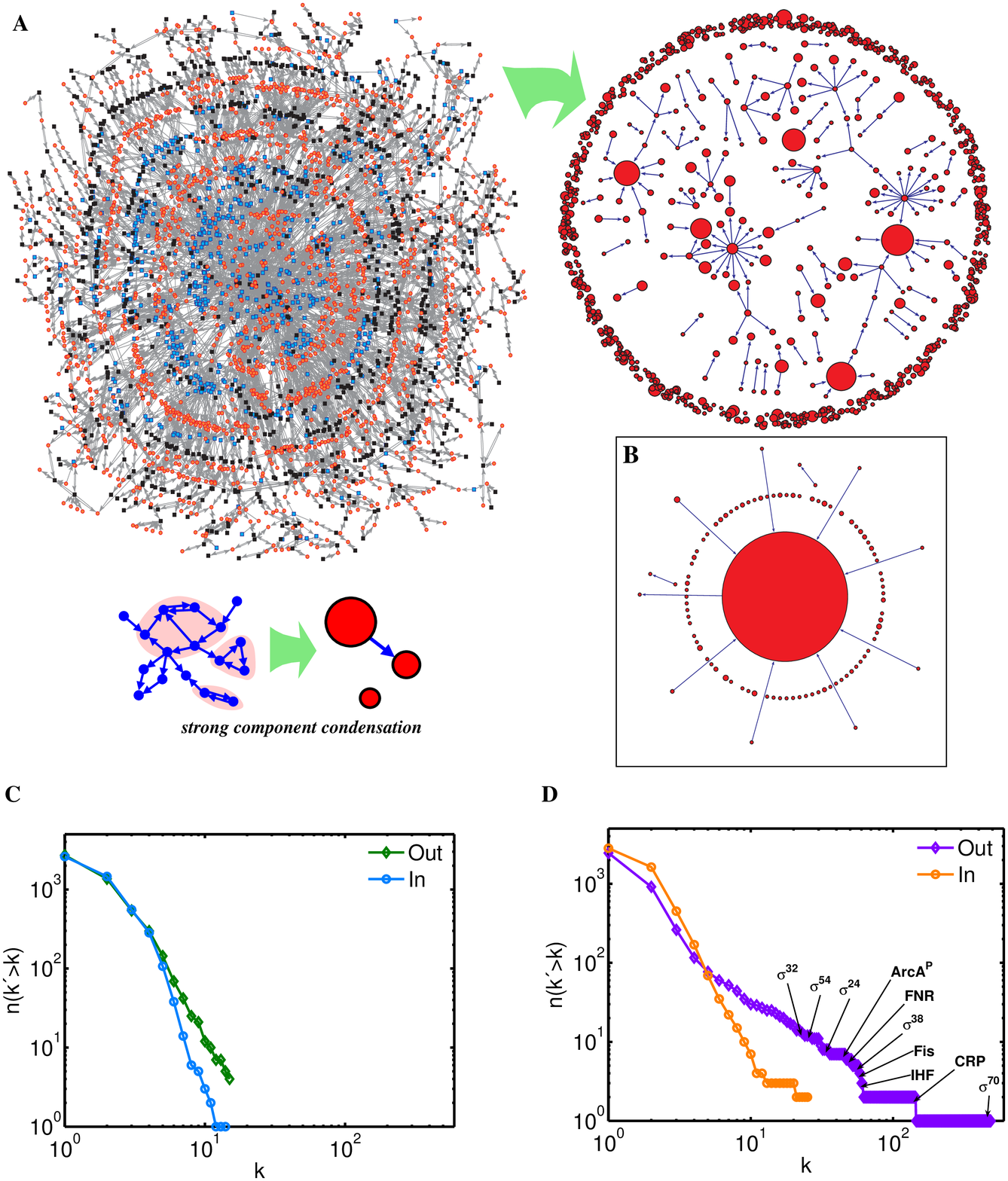}

\newpage

\textbf{Figure 2}\\

\includegraphics[height=.95\textheight]{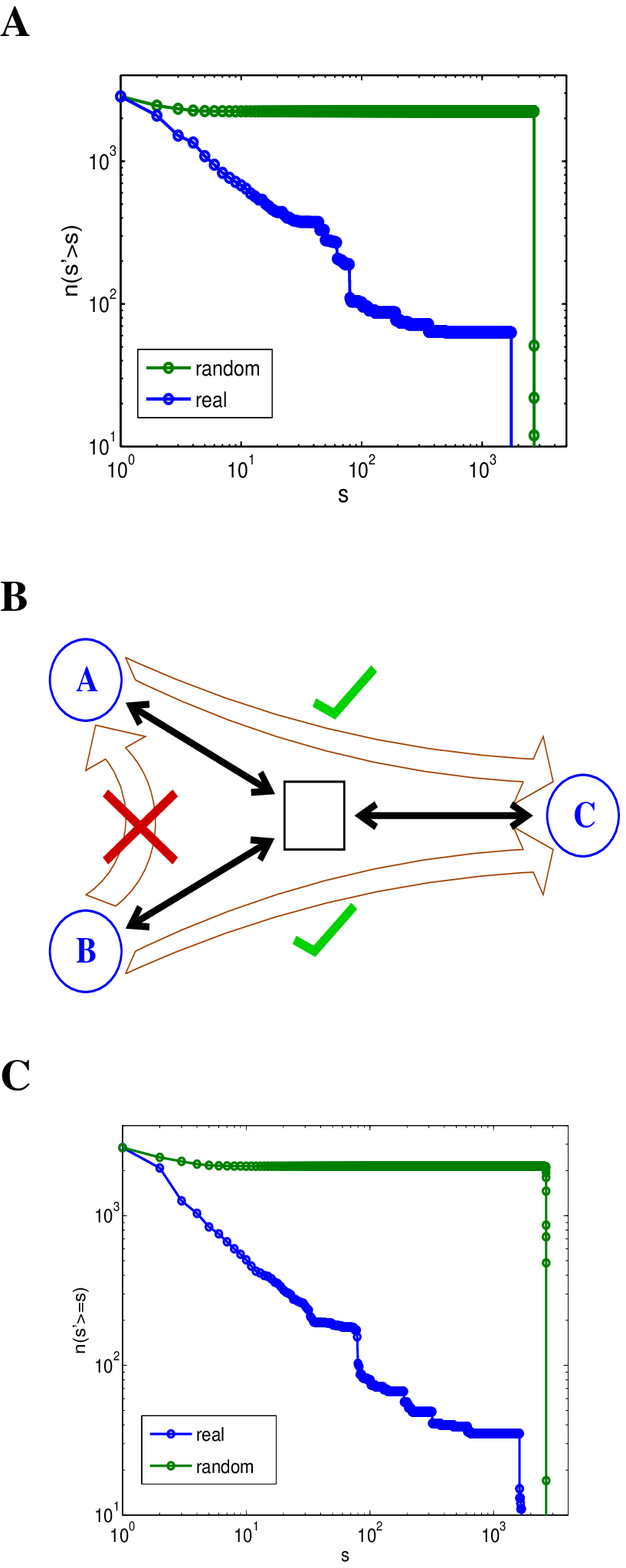}

\newpage

\textbf{Figure 3}\\

\includegraphics[height=.95\textheight]{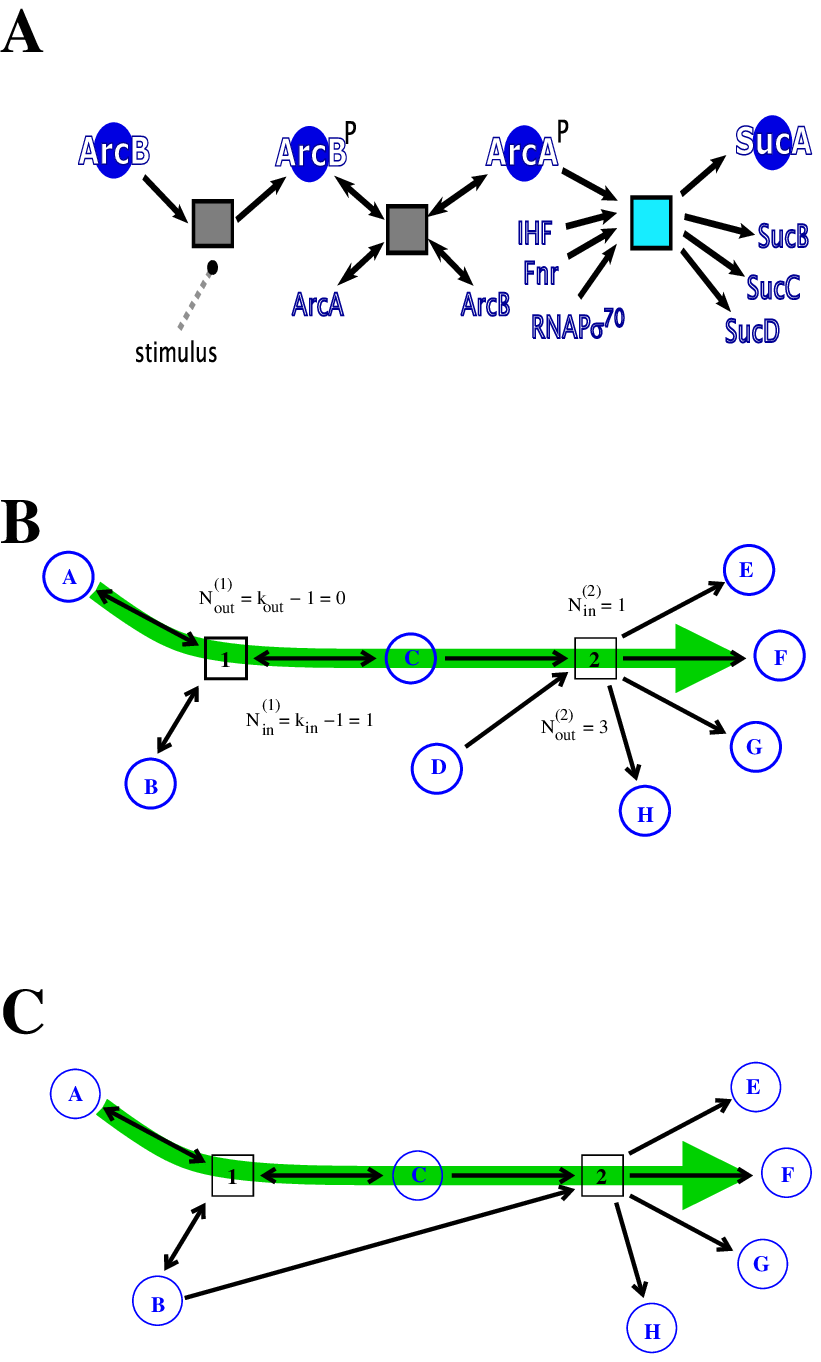}

\newpage

\textbf{Figure 4}\\

\includegraphics[height=.95\textheight]{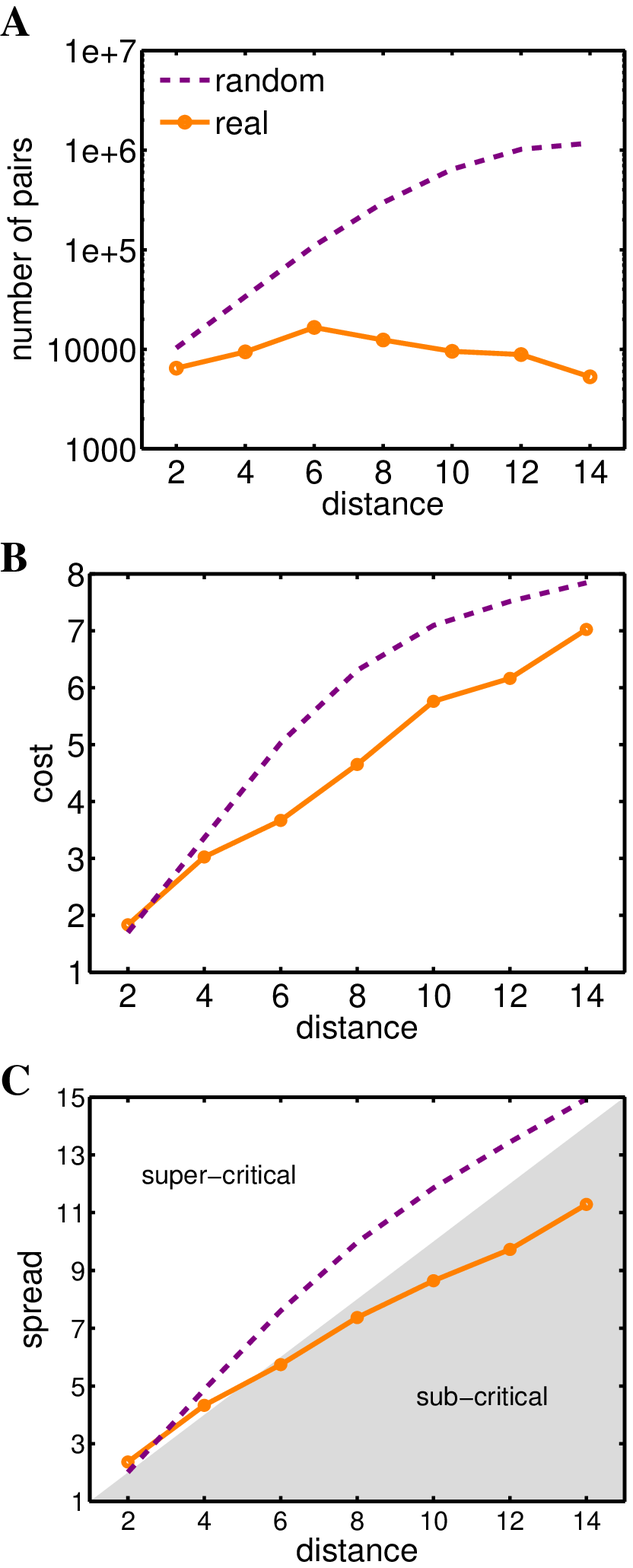}

\newpage

\textbf{Figure 5}\\

\includegraphics[width=\textwidth]{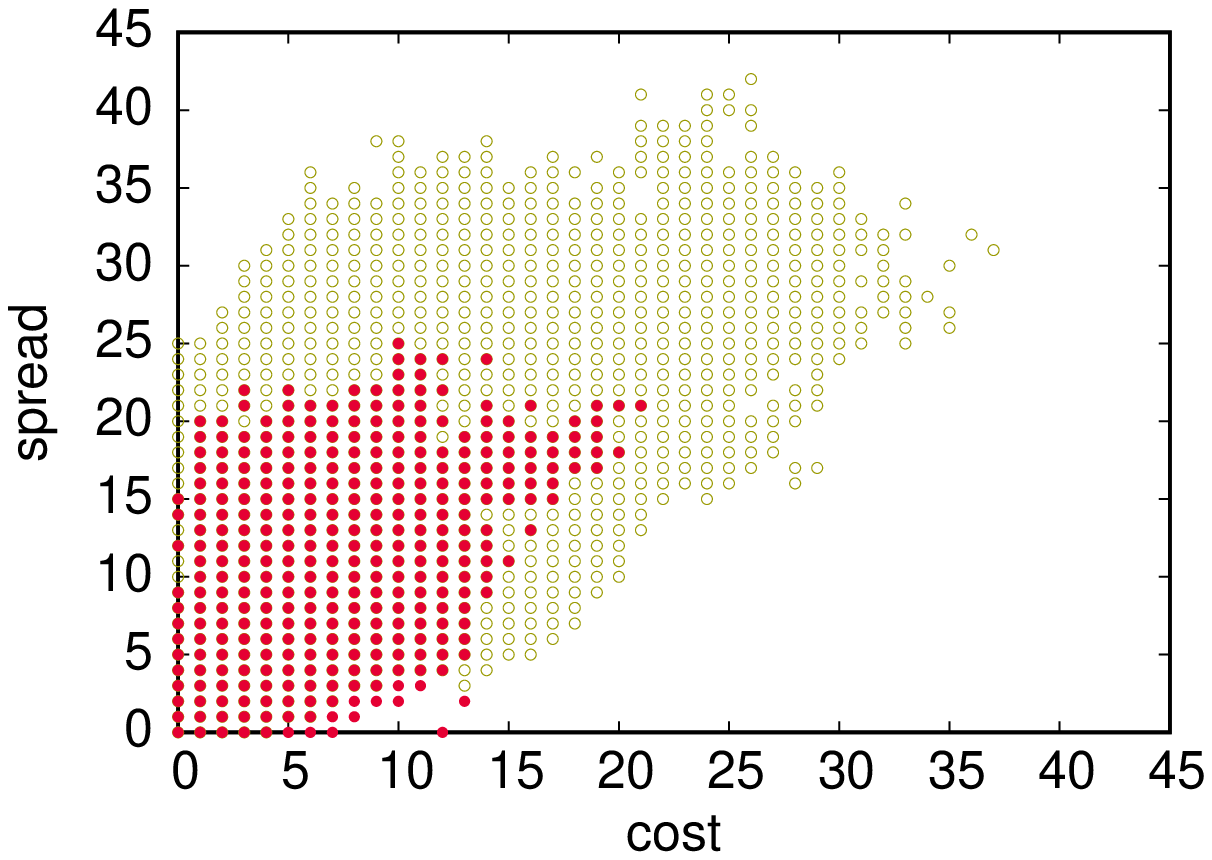}

\newpage

\textbf{Figure 6}\\

\includegraphics[height=.95\textheight]{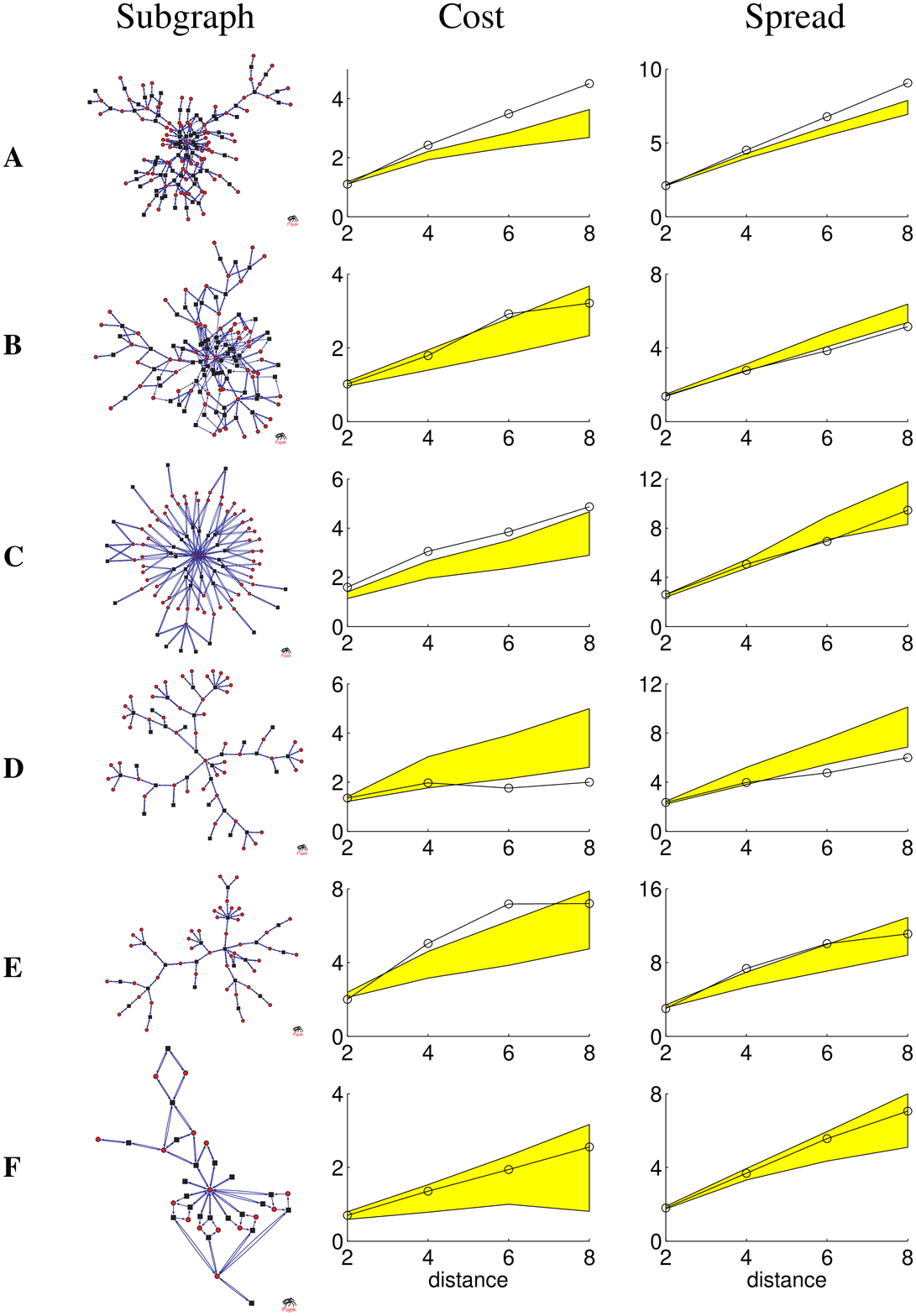}

\end{document}